\documentstyle[twoside,fleqn,espcrc2]{article}

\let\al=\alpha
\let\be=\beta
\let\ga=\gamma
\let\Ga=\Gamma
\let\de=\delta
\let\De=\Delta
\let\ep=\varepsilon
\let\eps=\epsilon
\let\ze=\zeta
\let\ka=\kappa
\let\la=\lambda
\let\La=\Lambda
\let\ph=\varphi
\let\del=\nabla
\let\si=\sigma
\let\Si=\Sigma
\let\th=\theta
\let\Up=\Upsilon
\let\om=\omega
\let\Om=\Omega
\let\Up=\Upsilon
\def\imp{~\Rightarrow}
\def\to{\rightarrow}
\let\p=\partial
\let\<=\langle
\let\>=\rangle
\let\ul=\underline
\let\txt=\textstyle
\let\dsp=\displaystyle
\let\h=\hbox
\let\ad=\dagger
\def\eqn#1{(\ref{#1})}  %Sticks parentheses around the LaTex "ref" macro
\def\Eqn#1{Eq.~(\ref{#1})}  %    includes ``Eq.'' in front
\def\e{ {\rm e} }
\def\pv{{\bf p}}
\def\beq{\begin{equation}}
\def\eeq{\end{equation}}
\def\ba{\begin{array}}
\def\bea{\begin{eqnarray}}
\def\ea{\end{array}}
\def\eea{\end{eqnarray}}

\def\lab{\label}
\def\slash{\!\!\!\!/\,}
\def\sslash{\!\!\!/\,}
\def\dotprod{\!\cdot\!}
\def\vfilll{\vskip 0pt plus 1filll}
\let\ol=\overline
\def\nl{\hfil\break}
\def\ee#1{ \times 10^{#1} }
\def\comment#1{ \hbox{[{\it Comment suppressed here.}\/]} }
\def\hide#1{}
\def\o{\over}      % TK: added
\def\O{ {\cal O} }
\def\tr{\hbox{tr}}
\def\Tr{\hbox{Tr}}
\def\P{\hbox{P}}
\def\hc{ {\rm h.c.} }
\def\Re{ {\rm Re}\, }
\def\ie{{\it i.e.}}
\def\eg{{\it e.g.}}
\def\comment#1{ }

\def\suN{su($N$)}
\def\SUN{SU($N$)}
\def\IR{\relax{\rm I\kern-.18em R}}
\def\IN{\relax{\rm I\kern-.18em N}}
\def\IB{\relax{\rm I\kern-.18em B}}
\def\IE{\relax{\rm I\kern-.18em E}}
\def\ZZ{\relax{\sf Z\kern-.4em Z}}     % use as $\ZZ$ or $\ZZ_p$
\def\IC{{\ontopss{$\scriptscriptstyle \mid$}{{\rm C}}{8.0}{0.415}}}
\def\ontopss#1#2#3#4{\raise#4ex \hbox{#1}\mkern-#3mu {#2}}

\def\lapp{ {\txt {{\txt <} \atop {\txt \sim}}} }
\def\gapp{ {\txt {{\txt >} \atop {\txt \sim}}} }
\newdimen\pmboffset
\def\oldpmb#1{\setbox0=\hbox{#1}%
 \copy0\kern-\wd0
 \kern\pmboffset\raise 1.732\pmboffset\copy0\kern-\wd0
 \kern\pmboffset\box0} 
\def\pmb#1{\mathchoice{\oldpmb{$\displaystyle#1$}}{\oldpmb{$\textstyle#1$}}
	{\oldpmb{$\scriptstyle#1$}}{\oldpmb{$\scriptscriptstyle#1$}}}

\def\appendix{\par                              % Have \appendix say
    \setcounter{section}{0}                     % `Appendix A', not just `A'
    \setcounter{subsection}{0}
    \renewcommand{\theequation}{\Alph{section}.\arabic{equation}}
    \renewcommand{\thesection}{Appendix \Alph{section}
                \setcounter{equation}{0}  } %Have eqns numbered (A.1) etc
}

\catcode`\@=11

\def\applabel#1{\@bsphack
  \protected@write\@auxout{}%
         {\string\newlabel{#1}{{\Alph{section}}{\thepage}}}%
  \@esphack}
\def\subsection{\@startsection{subsection}{2}{\z@}{-3.25ex plus -1ex minus 
 -.2ex}{1.5ex plus .2ex}{\normalsize\bf}}

\def\subsubsection{\@startsection{subsubsection}{3}{\z@}{-3.25ex plus
 -1ex minus -.2ex}{1.5ex plus .2ex}{\normalsize}}

\def\@eqnnum{%
\savebox{\eqnumb}{\rm (\theequation)}%
\settowidth{\numblen}{\usebox{\eqnumb}}%
\makebox[\numblen][l]{\usebox{\eqnumb}~~~\usebox{\eqlabel}}}

\catcode`\@=12

\newcommand{\skipover}[1]{}
\newcommand{\nn}{\nonumber \\}
\newcommand{\pl}{{\rm pl}}
\newcommand{\rt}{{\rm rt}}
\newcommand{\pg}{{\rm pg}}
\newcommand{\order}{{\cal O}}
\def\half {{\txt {1\over 2}}}
\def\third{{\txt {1\over 3}}}
\def\sixth{{\txt {1\over 6}}}

\def\Dsl{D\slash}
\def\Fmn{F_{\mu\nu}}
\def\Fcmn{F_{\! c \, \mu\nu}}
\def\pbar{{\bar p}}
\def\phat{\hat{p}}
\def\psibar{{\bar\psi}}
\def\Omb{{\bar\Om}}
\def\chibar{{\bar\chi}}
\def\={\!=\!}
\def\+{\,+\,}
\def\-{\,-\,}

\def\Csw{C_F}
\def\csw{c_{\rm sw}}
\def\MeV{{\rm MeV}}
\def\GeV{{\rm GeV}}
\def\fm{{\rm fm}}
\newsavebox{\eqlabel}

\catcode`\@=11
\newlength{\numblen}
\newsavebox{\eqnumb}
\def\@eqnnum{%
\savebox{\eqnumb}{\rm (\theequation)}%
\settowidth{\numblen}{\usebox{\eqnumb}}%
\makebox[\numblen][l]{\usebox{\eqnumb}~~~\usebox{\eqlabel}}%
}
\catcode`\@=12

\newenvironment{equationwithlabel}[1]{ %
   \[ \label{#1} } { \]}\savebox{\eqlabel}{~}
\newcommand{\beql}[1]{\begin{equationwithlabel}{#1}}
\newcommand{\eeql}{\end{equationwithlabel}}

\newenvironment{eqnarraywithlabel}[1]{ %
  \begin{eqnarray}\label{#1} }{\end{eqnarray}\savebox{\eqlabel}{~}}

\newcommand{\beal}[1]{\begin{eqnarraywithlabel}{#1}}
\newcommand{\eeal}{\end{eqnarraywithlabel}}
\newcommand{\ttbs}{\char'134}
\newcommand{\AmS}{{\protect\the\textfont2
  A\kern-.1667em\lower.5ex\hbox{M}\kern-.125emS}}

\title{Mean-link tadpole improvement of SW and D234 actions}

\author{Mark Alford\thanks{Presenter}\address{School of Natural Sciences,
        Institute for Advanced Study, 
	Princeton, NJ 08540, USA},
        Timothy R. Klassen\address{SCRI,  Florida State University,
        Tallahassee, FL 32306-4130, USA },
        and 
        Peter Lepage\address{Lab of Nuclear Studies, Cornell University,
        Ithaca, NY 14853, USA } 
}
       
\begin{document}

\begin{abstract}

We investigate a tadpole-improved  tree-level-$\O(a^3)$-accurate
action, D234c, on coarse lattices, using the mean link in Landau
gauge to measure the tadpole contribution.
We find that D234c shows much better rotational invariance than
Sheikholeslami-Wohlert, and that mean-link tadpole improvement
gives much better hadron mass scaling than plaquette tadpole
improvement, with finite-$a$ errors at $0.25~\fm$ of only a few percent.
% For D234c, the vector meson mass shows scaling violation
% of $2(1)\%$ at $a=0.25~\fm$, and $7(1)\%$ at $a=0.4~\fm$.

We explore the effects of possible $\O(\al_s)$ changes to the
improvement coefficients, and find that the two leading coefficients
can be independently tuned: hadron masses are only sensitive to the
clover coefficient $\Csw$, while hadron dispersion relations are only
sensitive to the third derivative coefficient $C_3$.  Preliminary
non-perturbative tuning of these coefficients yields values that are
consistent with the expected size of perturbative corrections.

\end{abstract}

\maketitle

\section{Introduction}
Because the cost of lattice QCD calculations is so sensitive to the
lattice spacing, improved actions on coarse lattices are an attractive
approach. In this paper we study a tadpole-improved 
(TI) tree-level $\O(a^3)$-accurate
action, D234c, comparing its predictions with those of the
Sheikholeslami-Wohlert (SW)  
tadpole-improved tree-level $\O(a)$-accurate action
\cite{SW}.
Since we are interested in finite-$a$ errors, we use 
quark masses near the strange quark mass, thereby avoiding
the separate complications of chiral extrapolation.

We use the mean link in Landau gauge rather than the traditional
plaquette prescription for calculating our TI factor
$u_0$. This is known to give smaller scaling errors in the NRQCD
charmonium hyperfine splitting \cite{Trottier}, and to give a clover
coefficient for Wilson glue that agrees more closely \cite{gplTsukuba} with the
non-perturbatively determined value \cite{alpha}.

Our glue action is a tree-level tadpole-improved plaquette and 
rectangle action \cite{LW,Alf1},
\beq
\label{glue}
 S = - \beta \sum_{x,\mu}
\left\{ 
\frac{5}{3}  \frac{P_{\mu\nu}(x)}{u_0^4}
- {R_{\mu\nu}(x) + R_{\nu\mu}(x) \over 12 u_0^6}
%- \frac{1}{12} \frac{R_{\mu\nu}(x)}{u_0^6}
%- \frac{1}{12} \frac{R_{\nu\mu}(x)}{u_0^6}
\right\}
\eeq
where $P_{\mu\nu}$ and $R_{\mu\nu}$ are the plaquette and $2\times 1$
rectangle.

For hadron spectrum measurements we generated gluon configurations at
two lattice spacings, $0.40~\fm$ and $0.25~\fm$, using lattices of
the same physical size ($2~\fm$) at both lattice spacings.
%(see Table \ref{tab:glue}).  
% The lattice spacing was determined by NRQCD charmonium simulations.
% using the experimental value of $458~\MeV$ for the spin-averaged $P-S$
% splitting. 

The D234c Dirac operator \cite{Improving} is
\[
\ba{l}
 M = \dsp m_c (1 + {r a m_c\over 2 }) 
 +\sum_\mu \Bigl\{
 \ga_\mu \De_\mu  - {C_3\over 6} a^2 \ga_\mu\De^{(3)}_\mu \Bigr. \nn
 + \dsp r \Bigl. \Bigl[  - 
 {1\over 2} a  \Delta^{(2)}_\mu
 - {\Csw\over 4} a \sum_{\nu} \sigma_{\mu\nu}  \Fmn
 + {C_4 \over 24} a^3 \De^{(4)}_\mu  \Bigr]\Bigr\}~.
\ea
\]
with $r\=C_F\=C_3\=C_4\=1$ at tree level.
$\De^{(n)}_\mu$ is a tadpole-improved lattice discretization of the 
gauge-covariant $n$'th derivative, and $\Fmn$
is the improved field strength \cite{Improving}.

\section{Hadron mass scaling}

\begin{figure}[htb]
\begin{center}
%\input{gnuplot/phi70.tex}
% GNUPLOT: LaTeX picture
\setlength{\unitlength}{0.240900pt}
\ifx\plotpoint\undefined\newsavebox{\plotpoint}\fi
\sbox{\plotpoint}{\rule[-0.200pt]{0.400pt}{0.400pt}}%
\begin{picture}(900,720)(0,0)
\font\gnuplot=cmr10 at 10pt
\gnuplot
\sbox{\plotpoint}{\rule[-0.200pt]{0.400pt}{0.400pt}}%
\put(176.0,68.0){\rule[-0.200pt]{0.400pt}{151.526pt}}
\put(176.0,78.0){\rule[-0.200pt]{4.818pt}{0.400pt}}
\put(154,78){\makebox(0,0)[r]{0.7}}
\put(816.0,78.0){\rule[-0.200pt]{4.818pt}{0.400pt}}
\put(176.0,173.0){\rule[-0.200pt]{4.818pt}{0.400pt}}
\put(154,173){\makebox(0,0)[r]{0.8}}
\put(816.0,173.0){\rule[-0.200pt]{4.818pt}{0.400pt}}
\put(176.0,268.0){\rule[-0.200pt]{4.818pt}{0.400pt}}
\put(154,268){\makebox(0,0)[r]{0.9}}
\put(816.0,268.0){\rule[-0.200pt]{4.818pt}{0.400pt}}
\put(176.0,363.0){\rule[-0.200pt]{4.818pt}{0.400pt}}
\put(154,363){\makebox(0,0)[r]{1}}
\put(816.0,363.0){\rule[-0.200pt]{4.818pt}{0.400pt}}
\put(176.0,459.0){\rule[-0.200pt]{4.818pt}{0.400pt}}
\put(154,459){\makebox(0,0)[r]{1.1}}
\put(816.0,459.0){\rule[-0.200pt]{4.818pt}{0.400pt}}
\put(176.0,554.0){\rule[-0.200pt]{4.818pt}{0.400pt}}
\put(154,554){\makebox(0,0)[r]{1.2}}
\put(816.0,554.0){\rule[-0.200pt]{4.818pt}{0.400pt}}
\put(176.0,649.0){\rule[-0.200pt]{4.818pt}{0.400pt}}
\put(154,649){\makebox(0,0)[r]{1.3}}
\put(816.0,649.0){\rule[-0.200pt]{4.818pt}{0.400pt}}
\put(176.0,68.0){\rule[-0.200pt]{0.400pt}{4.818pt}}
\put(176,23){\makebox(0,0){0}}
\put(176.0,677.0){\rule[-0.200pt]{0.400pt}{4.818pt}}
\put(506.0,68.0){\rule[-0.200pt]{0.400pt}{4.818pt}}
\put(506,23){\makebox(0,0){0.1}}
\put(506.0,677.0){\rule[-0.200pt]{0.400pt}{4.818pt}}
\put(836.0,68.0){\rule[-0.200pt]{0.400pt}{4.818pt}}
\put(836,23){\makebox(0,0){0.2}}
\put(836.0,677.0){\rule[-0.200pt]{0.400pt}{4.818pt}}
\put(176.0,68.0){\rule[-0.200pt]{158.994pt}{0.400pt}}
\put(836.0,68.0){\rule[-0.200pt]{0.400pt}{151.526pt}}
\put(176.0,697.0){\rule[-0.200pt]{158.994pt}{0.400pt}}
\put(11,392){\makebox(0,0)[l]{\shortstack{$\phi$ mass\\ \\ (GeV)}}}
\put(572,-17){\makebox(0,0)[l]{$a^2$ (fm${}^2$)}}
\put(176.0,68.0){\rule[-0.200pt]{0.400pt}{151.526pt}}
\put(706,632){\makebox(0,0)[r]{{\small D234c, mean-link TI}}}
\put(750,632){\circle*{18}}
\put(704,334){\circle*{18}}
\put(382,381){\circle*{18}}
\put(728.0,632.0){\rule[-0.200pt]{15.899pt}{0.400pt}}
\put(728.0,622.0){\rule[-0.200pt]{0.400pt}{4.818pt}}
\put(794.0,622.0){\rule[-0.200pt]{0.400pt}{4.818pt}}
\put(704.0,326.0){\rule[-0.200pt]{0.400pt}{3.854pt}}
\put(694.0,326.0){\rule[-0.200pt]{4.818pt}{0.400pt}}
\put(694.0,342.0){\rule[-0.200pt]{4.818pt}{0.400pt}}
\put(382.0,371.0){\rule[-0.200pt]{0.400pt}{4.577pt}}
\put(372.0,371.0){\rule[-0.200pt]{4.818pt}{0.400pt}}
\put(372.0,390.0){\rule[-0.200pt]{4.818pt}{0.400pt}}
\put(706,587){\makebox(0,0)[r]{{\small SW,    mean-link TI}}}
\put(750,587){\circle{18}}
\put(704,289){\circle{18}}
\put(382,403){\circle{18}}
\put(728.0,587.0){\rule[-0.200pt]{15.899pt}{0.400pt}}
\put(728.0,577.0){\rule[-0.200pt]{0.400pt}{4.818pt}}
\put(794.0,577.0){\rule[-0.200pt]{0.400pt}{4.818pt}}
\put(704.0,282.0){\rule[-0.200pt]{0.400pt}{3.373pt}}
\put(694.0,282.0){\rule[-0.200pt]{4.818pt}{0.400pt}}
\put(694.0,296.0){\rule[-0.200pt]{4.818pt}{0.400pt}}
\put(382.0,394.0){\rule[-0.200pt]{0.400pt}{4.577pt}}
\put(372.0,394.0){\rule[-0.200pt]{4.818pt}{0.400pt}}
\put(372.0,413.0){\rule[-0.200pt]{4.818pt}{0.400pt}}
\put(706,542){\makebox(0,0)[r]{{\small SW,  SCRI, plaq TI}}}
\put(750,542){\makebox(0,0){$\times$}}
\put(249,365){\makebox(0,0){$\times$}}
\put(271,358){\makebox(0,0){$\times$}}
\put(304,335){\makebox(0,0){$\times$}}
\put(370,299){\makebox(0,0){$\times$}}
\put(509,234){\makebox(0,0){$\times$}}
\put(692,144){\makebox(0,0){$\times$}}
\put(728.0,542.0){\rule[-0.200pt]{15.899pt}{0.400pt}}
\put(728.0,532.0){\rule[-0.200pt]{0.400pt}{4.818pt}}
\put(794.0,532.0){\rule[-0.200pt]{0.400pt}{4.818pt}}
\put(249.0,354.0){\rule[-0.200pt]{0.400pt}{5.541pt}}
\put(239.0,354.0){\rule[-0.200pt]{4.818pt}{0.400pt}}
\put(239.0,377.0){\rule[-0.200pt]{4.818pt}{0.400pt}}
\put(271.0,345.0){\rule[-0.200pt]{0.400pt}{6.022pt}}
\put(261.0,345.0){\rule[-0.200pt]{4.818pt}{0.400pt}}
\put(261.0,370.0){\rule[-0.200pt]{4.818pt}{0.400pt}}
\put(304.0,328.0){\rule[-0.200pt]{0.400pt}{3.373pt}}
\put(294.0,328.0){\rule[-0.200pt]{4.818pt}{0.400pt}}
\put(294.0,342.0){\rule[-0.200pt]{4.818pt}{0.400pt}}
\put(370.0,294.0){\rule[-0.200pt]{0.400pt}{2.409pt}}
\put(360.0,294.0){\rule[-0.200pt]{4.818pt}{0.400pt}}
\put(360.0,304.0){\rule[-0.200pt]{4.818pt}{0.400pt}}
\put(509.0,228.0){\rule[-0.200pt]{0.400pt}{2.891pt}}
\put(499.0,228.0){\rule[-0.200pt]{4.818pt}{0.400pt}}
\put(499.0,240.0){\rule[-0.200pt]{4.818pt}{0.400pt}}
\put(692.0,141.0){\rule[-0.200pt]{0.400pt}{1.445pt}}
\put(682.0,141.0){\rule[-0.200pt]{4.818pt}{0.400pt}}
\put(682.0,147.0){\rule[-0.200pt]{4.818pt}{0.400pt}}
\put(706,497){\makebox(0,0)[r]{{\small Wil, SCRI, plaq TI}}}
\put(750,497){\makebox(0,0){$+$}}
\put(249,178){\makebox(0,0){$+$}}
\put(271,157){\makebox(0,0){$+$}}
\put(304,112){\makebox(0,0){$+$}}
\put(728.0,497.0){\rule[-0.200pt]{15.899pt}{0.400pt}}
\put(728.0,487.0){\rule[-0.200pt]{0.400pt}{4.818pt}}
\put(794.0,487.0){\rule[-0.200pt]{0.400pt}{4.818pt}}
\put(249.0,167.0){\rule[-0.200pt]{0.400pt}{5.059pt}}
\put(239.0,167.0){\rule[-0.200pt]{4.818pt}{0.400pt}}
\put(239.0,188.0){\rule[-0.200pt]{4.818pt}{0.400pt}}
\put(271.0,148.0){\rule[-0.200pt]{0.400pt}{4.095pt}}
\put(261.0,148.0){\rule[-0.200pt]{4.818pt}{0.400pt}}
\put(261.0,165.0){\rule[-0.200pt]{4.818pt}{0.400pt}}
\put(304.0,107.0){\rule[-0.200pt]{0.400pt}{2.650pt}}
\put(294.0,107.0){\rule[-0.200pt]{4.818pt}{0.400pt}}
\put(294.0,118.0){\rule[-0.200pt]{4.818pt}{0.400pt}}
\sbox{\plotpoint}{\rule[-0.500pt]{1.000pt}{1.000pt}}%
\put(176,397){\usebox{\plotpoint}}
\put(176.00,397.00){\usebox{\plotpoint}}
\multiput(183,394)(18.564,-9.282){0}{\usebox{\plotpoint}}
\put(194.58,387.81){\usebox{\plotpoint}}
\multiput(196,387)(19.077,-8.176){0}{\usebox{\plotpoint}}
\multiput(203,384)(18.564,-9.282){0}{\usebox{\plotpoint}}
\put(213.41,379.11){\usebox{\plotpoint}}
\multiput(216,378)(18.021,-10.298){0}{\usebox{\plotpoint}}
\multiput(223,374)(18.564,-9.282){0}{\usebox{\plotpoint}}
\put(231.91,369.75){\usebox{\plotpoint}}
\multiput(236,368)(19.077,-8.176){0}{\usebox{\plotpoint}}
\multiput(243,365)(17.270,-11.513){0}{\usebox{\plotpoint}}
\put(250.36,360.42){\usebox{\plotpoint}}
\multiput(256,358)(19.077,-8.176){0}{\usebox{\plotpoint}}
\put(268.83,351.11){\usebox{\plotpoint}}
\multiput(269,351)(19.077,-8.176){0}{\usebox{\plotpoint}}
\multiput(276,348)(19.077,-8.176){0}{\usebox{\plotpoint}}
\put(287.76,342.62){\usebox{\plotpoint}}
\multiput(289,342)(18.021,-10.298){0}{\usebox{\plotpoint}}
\multiput(296,338)(19.077,-8.176){0}{\usebox{\plotpoint}}
\put(306.30,333.35){\usebox{\plotpoint}}
\multiput(309,332)(19.077,-8.176){0}{\usebox{\plotpoint}}
\multiput(316,329)(18.021,-10.298){0}{\usebox{\plotpoint}}
\put(324.84,324.08){\usebox{\plotpoint}}
\multiput(329,322)(19.077,-8.176){0}{\usebox{\plotpoint}}
\multiput(336,319)(18.021,-10.298){0}{\usebox{\plotpoint}}
\put(343.38,314.81){\usebox{\plotpoint}}
\multiput(349,312)(19.077,-8.176){0}{\usebox{\plotpoint}}
\put(362.30,306.30){\usebox{\plotpoint}}
\multiput(363,306)(17.270,-11.513){0}{\usebox{\plotpoint}}
\multiput(369,302)(19.077,-8.176){0}{\usebox{\plotpoint}}
\put(380.75,296.96){\usebox{\plotpoint}}
\multiput(383,296)(18.564,-9.282){0}{\usebox{\plotpoint}}
\multiput(389,293)(18.021,-10.298){0}{\usebox{\plotpoint}}
\put(399.25,287.61){\usebox{\plotpoint}}
\multiput(403,286)(18.564,-9.282){0}{\usebox{\plotpoint}}
\multiput(409,283)(18.021,-10.298){0}{\usebox{\plotpoint}}
\put(417.76,278.25){\usebox{\plotpoint}}
\multiput(423,276)(18.564,-9.282){0}{\usebox{\plotpoint}}
\multiput(429,273)(19.077,-8.176){0}{\usebox{\plotpoint}}
\put(436.63,269.64){\usebox{\plotpoint}}
\multiput(443,266)(18.564,-9.282){0}{\usebox{\plotpoint}}
\put(455.17,260.36){\usebox{\plotpoint}}
\multiput(456,260)(18.021,-10.298){0}{\usebox{\plotpoint}}
\multiput(463,256)(18.564,-9.282){0}{\usebox{\plotpoint}}
\put(473.67,251.00){\usebox{\plotpoint}}
\multiput(476,250)(19.077,-8.176){0}{\usebox{\plotpoint}}
\multiput(483,247)(17.270,-11.513){0}{\usebox{\plotpoint}}
\put(492.12,241.66){\usebox{\plotpoint}}
\multiput(496,240)(19.077,-8.176){0}{\usebox{\plotpoint}}
\multiput(503,237)(18.564,-9.282){0}{\usebox{\plotpoint}}
\put(510.92,232.90){\usebox{\plotpoint}}
\multiput(516,230)(19.077,-8.176){0}{\usebox{\plotpoint}}
\multiput(523,227)(18.564,-9.282){0}{\usebox{\plotpoint}}
\put(529.50,223.71){\usebox{\plotpoint}}
\multiput(536,220)(19.077,-8.176){0}{\usebox{\plotpoint}}
\put(548.06,214.47){\usebox{\plotpoint}}
\multiput(549,214)(19.077,-8.176){0}{\usebox{\plotpoint}}
\multiput(556,211)(18.021,-10.298){0}{\usebox{\plotpoint}}
\put(566.60,205.20){\usebox{\plotpoint}}
\multiput(569,204)(19.077,-8.176){0}{\usebox{\plotpoint}}
\multiput(576,201)(19.077,-8.176){0}{\usebox{\plotpoint}}
\put(585.36,196.42){\usebox{\plotpoint}}
\multiput(589,194)(19.077,-8.176){0}{\usebox{\plotpoint}}
\multiput(596,191)(19.077,-8.176){0}{\usebox{\plotpoint}}
\put(603.96,187.36){\usebox{\plotpoint}}
\multiput(609,184)(19.077,-8.176){0}{\usebox{\plotpoint}}
\put(622.51,178.21){\usebox{\plotpoint}}
\multiput(623,178)(18.564,-9.282){0}{\usebox{\plotpoint}}
\multiput(629,175)(18.021,-10.298){0}{\usebox{\plotpoint}}
\put(641.01,168.85){\usebox{\plotpoint}}
\multiput(643,168)(18.564,-9.282){0}{\usebox{\plotpoint}}
\multiput(649,165)(19.077,-8.176){0}{\usebox{\plotpoint}}
\put(659.71,159.88){\usebox{\plotpoint}}
\multiput(663,158)(18.564,-9.282){0}{\usebox{\plotpoint}}
\multiput(669,155)(19.077,-8.176){0}{\usebox{\plotpoint}}
\put(678.29,150.69){\usebox{\plotpoint}}
\multiput(683,148)(18.564,-9.282){0}{\usebox{\plotpoint}}
\multiput(689,145)(19.077,-8.176){0}{\usebox{\plotpoint}}
\put(696.92,141.60){\usebox{\plotpoint}}
\multiput(703,139)(17.270,-11.513){0}{\usebox{\plotpoint}}
\put(715.37,132.27){\usebox{\plotpoint}}
\multiput(716,132)(19.077,-8.176){0}{\usebox{\plotpoint}}
\multiput(723,129)(18.564,-9.282){0}{\usebox{\plotpoint}}
\put(733.99,123.15){\usebox{\plotpoint}}
\multiput(736,122)(19.077,-8.176){0}{\usebox{\plotpoint}}
\multiput(743,119)(18.564,-9.282){0}{\usebox{\plotpoint}}
\put(752.58,113.96){\usebox{\plotpoint}}
\multiput(756,112)(19.077,-8.176){0}{\usebox{\plotpoint}}
\multiput(763,109)(18.564,-9.282){0}{\usebox{\plotpoint}}
\put(771.29,105.02){\usebox{\plotpoint}}
\multiput(776,103)(18.021,-10.298){0}{\usebox{\plotpoint}}
\multiput(783,99)(18.564,-9.282){0}{\usebox{\plotpoint}}
\put(789.79,95.66){\usebox{\plotpoint}}
\multiput(796,93)(19.077,-8.176){0}{\usebox{\plotpoint}}
\put(808.31,86.46){\usebox{\plotpoint}}
\multiput(809,86)(19.077,-8.176){0}{\usebox{\plotpoint}}
\multiput(816,83)(19.077,-8.176){0}{\usebox{\plotpoint}}
\put(826.91,77.40){\usebox{\plotpoint}}
\multiput(829,76)(19.077,-8.176){0}{\usebox{\plotpoint}}
\put(836,73){\usebox{\plotpoint}}
\put(176,396){\usebox{\plotpoint}}
\multiput(176,396)(2.065,-20.652){4}{\usebox{\plotpoint}}
\put(185.66,313.61){\usebox{\plotpoint}}
\put(190.52,293.45){\usebox{\plotpoint}}
\put(197.24,273.82){\usebox{\plotpoint}}
\put(204.80,254.49){\usebox{\plotpoint}}
\put(213.48,235.68){\usebox{\plotpoint}}
\multiput(216,231)(9.840,-18.275){0}{\usebox{\plotpoint}}
\put(223.33,217.40){\usebox{\plotpoint}}
\put(234.11,199.70){\usebox{\plotpoint}}
\multiput(236,197)(11.902,-17.004){0}{\usebox{\plotpoint}}
\put(245.70,182.50){\usebox{\plotpoint}}
\multiput(249,177)(13.668,-15.620){0}{\usebox{\plotpoint}}
\put(258.28,166.07){\usebox{\plotpoint}}
\multiput(263,160)(12.453,-16.604){0}{\usebox{\plotpoint}}
\put(271.17,149.83){\usebox{\plotpoint}}
\multiput(276,145)(13.668,-15.620){0}{\usebox{\plotpoint}}
\put(285.14,134.50){\usebox{\plotpoint}}
\multiput(289,130)(14.676,-14.676){0}{\usebox{\plotpoint}}
\put(299.74,119.79){\usebox{\plotpoint}}
\multiput(303,117)(14.676,-14.676){0}{\usebox{\plotpoint}}
\put(314.64,105.36){\usebox{\plotpoint}}
\multiput(316,104)(15.759,-13.508){0}{\usebox{\plotpoint}}
\multiput(323,98)(15.945,-13.287){0}{\usebox{\plotpoint}}
\put(330.37,91.83){\usebox{\plotpoint}}
\multiput(336,87)(16.889,-12.064){0}{\usebox{\plotpoint}}
\put(346.35,78.65){\usebox{\plotpoint}}
\multiput(349,76)(16.889,-12.064){0}{\usebox{\plotpoint}}
\multiput(356,71)(16.604,-12.453){0}{\usebox{\plotpoint}}
\put(360,68){\usebox{\plotpoint}}
\end{picture}

\end{center}
\vspace{-3ex}
\caption{ 
Mass of $\phi$ meson as a function of lattice spacing, $P/V=0.70$.
Mean link TI ($\bullet$ D234, $\circ$ SW) is more continuum-like than 
plaquette TI ($\times$ SW, + Wilson, data supplied by SCRI)}
\label{fig:phi70}
\end{figure}
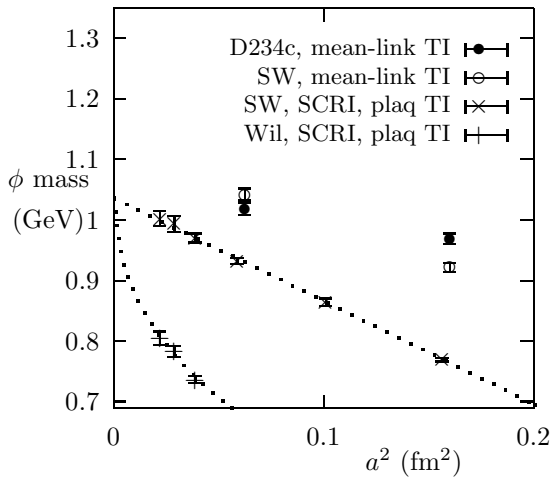

In Figure~\ref{fig:phi70} we show that mean-link tadpole improvement
leads to much smaller finite-$a$ errors for the vector meson mass.
(We fixed our lattice spacing relative to SCRI's by the
``Galilean quarkonium'' method \cite{longversion}.)
The remaining finite-$a$ errors
% The D234c results show finite-$a$ errors of 2(1)\% at 0.25\,fm and
% 7(1)\% at 0.4\,fm.  These 
are due to a mixture of radiative
corrections to the tree-level coupling constants, and
higher-order interactions not included in the D234c action.

To investigate the sensitivity of the hadron masses to radiative
corrections we measured the fractional change caused by multiplying
each coupling constant in turn by $1+\alpha_s$,
(Table~\ref{tab:sensitivity}). We see that the only coupling for which
radiative corrections are important is the clover
coupling,~$C_F$ \cite{Tim}.  
Mean-link tadpole improvement increases the clover 
coefficient, and reduces finite-$a$ errors
in the vector mass for both D234c and SW,
as seen in Figure~\ref{fig:phi70}. The superiority of D234c will become
apparent when observables sensitive to Lorentz violation are
considered (see below).

We performed a quadratic $a\to 0$ extrapolation of SCRI's SW results
to estimate the continuum $V$ mass. Adjusting the clover
coefficient to give this $V$~mass at $a=0.25~\fm$, we find that choosing
\beq\label{CF}
C_F = 1 + \alpha_s/6,
\eeq
where $\al_s \equiv {3 / 2\pi\beta}$ is the bare coupling, reduces
the $V$ mass error at 0.25~\fm\ from 2(1)\% 
to 0(1)\%, and also at 0.4~\fm\ from 7(1)\% to 0(1)\%.  
This suggests that perturbative corrections to
$\Csw$ are relatively small after tadpole improvement. 
(The quadratic fit of SCRI's SW data may be too naive: the
true continuum value could be different by a few percent,
changing the coefficient of $\al_s$, but leaving it perturbative
in size.)

A perturbative
analysis of $\Csw$ through $\O(\al_s)$ will soon be completed
\cite{TrotLep}, and will be compared with this nonperturbative
estimate.  (Tuning of $C_F$ using PCAC is another
possibility \cite{TimRob}.)  Whatever the result of this comparison,
the non-perturbative formula \eqn{CF} should give hadron masses accurate to
a few percent for lattice spacings up to 0.4~\fm.

\def\st{\rule[-1.5ex]{0em}{4ex}} % Strut to give correct row spacing in tables
\begin{table}[hbt]
% space before first and after last column: 1.5pc
% space between columns: 3.0pc (twice the above)
%\setlength{\tabcolsep}{1.5pc}
% -----------------------------------------------------
% adapted from TeX book, p. 241
\newlength{\digitwidth} \settowidth{\digitwidth}{\rm 0}
\catcode`?=\active \def?{\kern\digitwidth}
% -----------------------------------------------------
%\caption{Biologically treated effluents (mg/l)}
%\begin{tabular*}{\textwidth}{llllll}
\begin{tabular}{llllll}
\hline
\st  & \multicolumn{2}{c}{$\Csw$} & $C_3$ & $C_4$  \\
\st $a$ (fm) & 0.25 & 0.4             & 0.4   & 0.4  \\
\hline
\st P & 11(1) & 35(1)  & 0.9(1) & -3.2(1)   \\
\st V & 10(1) & 35(1)  & 1.0(2) & -3.0(3)   \\
\st N & 6(2)  & 26(2)  & 0.9(3) & -2.7(4)   \\
\st D & 8(1)  & 25(2)  & 1.0(1) & -2.1(4)   \\
\hline
\vspace{-0.5ex}
\end{tabular}
\caption{
Percentage change in hadron masses at $P/V\approx 0.76$
when individual coefficients in the mean-link tree-level TI
D234c quark action  are multiplied by $1+\alpha_s$.
}
\label{tab:sensitivity}
\vspace{-2ex}
\end{table}

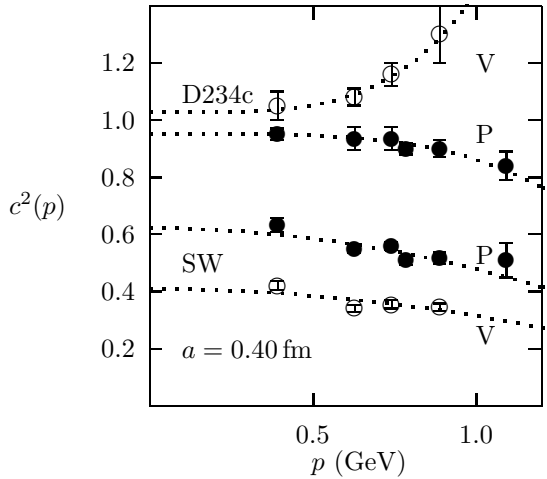
\begin{figure}[htb]
\begin{center}
%\input{gnuplot/csq.tex}
% GNUPLOT: LaTeX picture
\setlength{\unitlength}{0.240900pt}
\ifx\plotpoint\undefined\newsavebox{\plotpoint}\fi
\sbox{\plotpoint}{\rule[-0.200pt]{0.400pt}{0.400pt}}%
\begin{picture}(900,765)(0,0)
\font\gnuplot=cmr10 at 10pt
\gnuplot
\sbox{\plotpoint}{\rule[-0.200pt]{0.400pt}{0.400pt}}%
\put(220.0,113.0){\rule[-0.200pt]{148.394pt}{0.400pt}}
\put(220.0,113.0){\rule[-0.200pt]{0.400pt}{151.526pt}}
\put(220.0,203.0){\rule[-0.200pt]{4.818pt}{0.400pt}}
\put(198,203){\makebox(0,0)[r]{0.2}}
\put(816.0,203.0){\rule[-0.200pt]{4.818pt}{0.400pt}}
\put(220.0,293.0){\rule[-0.200pt]{4.818pt}{0.400pt}}
\put(198,293){\makebox(0,0)[r]{0.4}}
\put(816.0,293.0){\rule[-0.200pt]{4.818pt}{0.400pt}}
\put(220.0,383.0){\rule[-0.200pt]{4.818pt}{0.400pt}}
\put(198,383){\makebox(0,0)[r]{0.6}}
\put(816.0,383.0){\rule[-0.200pt]{4.818pt}{0.400pt}}
\put(220.0,472.0){\rule[-0.200pt]{4.818pt}{0.400pt}}
\put(198,472){\makebox(0,0)[r]{0.8}}
\put(816.0,472.0){\rule[-0.200pt]{4.818pt}{0.400pt}}
\put(220.0,562.0){\rule[-0.200pt]{4.818pt}{0.400pt}}
\put(198,562){\makebox(0,0)[r]{1.0}}
\put(816.0,562.0){\rule[-0.200pt]{4.818pt}{0.400pt}}
\put(220.0,652.0){\rule[-0.200pt]{4.818pt}{0.400pt}}
\put(198,652){\makebox(0,0)[r]{1.2}}
\put(816.0,652.0){\rule[-0.200pt]{4.818pt}{0.400pt}}
\put(477.0,113.0){\rule[-0.200pt]{0.400pt}{4.818pt}}
\put(477,68){\makebox(0,0){0.5}}
\put(477.0,722.0){\rule[-0.200pt]{0.400pt}{4.818pt}}
\put(733.0,113.0){\rule[-0.200pt]{0.400pt}{4.818pt}}
\put(733,68){\makebox(0,0){1.0}}
\put(733.0,722.0){\rule[-0.200pt]{0.400pt}{4.818pt}}
\put(220.0,113.0){\rule[-0.200pt]{148.394pt}{0.400pt}}
\put(836.0,113.0){\rule[-0.200pt]{0.400pt}{151.526pt}}
\put(220.0,742.0){\rule[-0.200pt]{148.394pt}{0.400pt}}
\put(45,427){\makebox(0,0){$c^2(p)$}}
\put(550,23){\makebox(0,0){$p$ (GeV)}}
\put(271,603){\makebox(0,0)[l]{D234c}}
\put(271,338){\makebox(0,0)[l]{SW}}
\put(271,203){\makebox(0,0)[l]{$a=0.40$\,fm}}
\put(733,540){\makebox(0,0)[l]{P}}
\put(733,652){\makebox(0,0)[l]{V}}
\put(733,351){\makebox(0,0)[l]{P}}
\put(733,225){\makebox(0,0)[l]{V}}
\put(220.0,113.0){\rule[-0.200pt]{0.400pt}{151.526pt}}
\put(421,540){\circle*{24}}
\put(542,533){\circle*{24}}
\put(600,533){\circle*{24}}
\put(623,517){\circle*{24}}
\put(676,517){\circle*{24}}
\put(780,490){\circle*{24}}
\put(421.0,531.0){\rule[-0.200pt]{0.400pt}{4.336pt}}
\put(411.0,531.0){\rule[-0.200pt]{4.818pt}{0.400pt}}
\put(411.0,549.0){\rule[-0.200pt]{4.818pt}{0.400pt}}
\put(542.0,515.0){\rule[-0.200pt]{0.400pt}{8.672pt}}
\put(532.0,515.0){\rule[-0.200pt]{4.818pt}{0.400pt}}
\put(532.0,551.0){\rule[-0.200pt]{4.818pt}{0.400pt}}
\put(600.0,515.0){\rule[-0.200pt]{0.400pt}{8.672pt}}
\put(590.0,515.0){\rule[-0.200pt]{4.818pt}{0.400pt}}
\put(590.0,551.0){\rule[-0.200pt]{4.818pt}{0.400pt}}
\put(623.0,508.0){\rule[-0.200pt]{0.400pt}{4.336pt}}
\put(613.0,508.0){\rule[-0.200pt]{4.818pt}{0.400pt}}
\put(613.0,526.0){\rule[-0.200pt]{4.818pt}{0.400pt}}
\put(676.0,504.0){\rule[-0.200pt]{0.400pt}{6.504pt}}
\put(666.0,504.0){\rule[-0.200pt]{4.818pt}{0.400pt}}
\put(666.0,531.0){\rule[-0.200pt]{4.818pt}{0.400pt}}
\put(780.0,468.0){\rule[-0.200pt]{0.400pt}{10.840pt}}
\put(770.0,468.0){\rule[-0.200pt]{4.818pt}{0.400pt}}
\put(770.0,513.0){\rule[-0.200pt]{4.818pt}{0.400pt}}
\put(421,585){\circle{24}}
\put(542,598){\circle{24}}
\put(600,634){\circle{24}}
\put(676,697){\circle{24}}
\put(421.0,562.0){\rule[-0.200pt]{0.400pt}{10.840pt}}
\put(411.0,562.0){\rule[-0.200pt]{4.818pt}{0.400pt}}
\put(411.0,607.0){\rule[-0.200pt]{4.818pt}{0.400pt}}
\put(542.0,585.0){\rule[-0.200pt]{0.400pt}{6.504pt}}
\put(532.0,585.0){\rule[-0.200pt]{4.818pt}{0.400pt}}
\put(532.0,612.0){\rule[-0.200pt]{4.818pt}{0.400pt}}
\put(600.0,616.0){\rule[-0.200pt]{0.400pt}{8.672pt}}
\put(590.0,616.0){\rule[-0.200pt]{4.818pt}{0.400pt}}
\put(590.0,652.0){\rule[-0.200pt]{4.818pt}{0.400pt}}
\put(676.0,652.0){\rule[-0.200pt]{0.400pt}{21.681pt}}
\put(666.0,652.0){\rule[-0.200pt]{4.818pt}{0.400pt}}
\put(666.0,742.0){\rule[-0.200pt]{4.818pt}{0.400pt}}
\put(421,397){\circle*{24}}
\put(542,360){\circle*{24}}
\put(600,365){\circle*{24}}
\put(623,342){\circle*{24}}
\put(676,345){\circle*{24}}
\put(780,342){\circle*{24}}
\put(421.0,386.0){\rule[-0.200pt]{0.400pt}{5.300pt}}
\put(411.0,386.0){\rule[-0.200pt]{4.818pt}{0.400pt}}
\put(411.0,408.0){\rule[-0.200pt]{4.818pt}{0.400pt}}
\put(542.0,354.0){\rule[-0.200pt]{0.400pt}{2.891pt}}
\put(532.0,354.0){\rule[-0.200pt]{4.818pt}{0.400pt}}
\put(532.0,366.0){\rule[-0.200pt]{4.818pt}{0.400pt}}
\put(600.0,358.0){\rule[-0.200pt]{0.400pt}{3.132pt}}
\put(590.0,358.0){\rule[-0.200pt]{4.818pt}{0.400pt}}
\put(590.0,371.0){\rule[-0.200pt]{4.818pt}{0.400pt}}
\put(623.0,335.0){\rule[-0.200pt]{0.400pt}{3.373pt}}
\put(613.0,335.0){\rule[-0.200pt]{4.818pt}{0.400pt}}
\put(613.0,349.0){\rule[-0.200pt]{4.818pt}{0.400pt}}
\put(676.0,336.0){\rule[-0.200pt]{0.400pt}{4.336pt}}
\put(666.0,336.0){\rule[-0.200pt]{4.818pt}{0.400pt}}
\put(666.0,354.0){\rule[-0.200pt]{4.818pt}{0.400pt}}
\put(780.0,315.0){\rule[-0.200pt]{0.400pt}{13.009pt}}
\put(770.0,315.0){\rule[-0.200pt]{4.818pt}{0.400pt}}
\put(770.0,369.0){\rule[-0.200pt]{4.818pt}{0.400pt}}
\put(421,302){\circle{24}}
\put(542,267){\circle{24}}
\put(600,272){\circle{24}}
\put(676,268){\circle{24}}
\put(421.0,295.0){\rule[-0.200pt]{0.400pt}{3.373pt}}
\put(411.0,295.0){\rule[-0.200pt]{4.818pt}{0.400pt}}
\put(411.0,309.0){\rule[-0.200pt]{4.818pt}{0.400pt}}
\put(542.0,261.0){\rule[-0.200pt]{0.400pt}{2.650pt}}
\put(532.0,261.0){\rule[-0.200pt]{4.818pt}{0.400pt}}
\put(532.0,272.0){\rule[-0.200pt]{4.818pt}{0.400pt}}
\put(600.0,266.0){\rule[-0.200pt]{0.400pt}{3.132pt}}
\put(590.0,266.0){\rule[-0.200pt]{4.818pt}{0.400pt}}
\put(590.0,279.0){\rule[-0.200pt]{4.818pt}{0.400pt}}
\put(676.0,262.0){\rule[-0.200pt]{0.400pt}{2.891pt}}
\put(666.0,262.0){\rule[-0.200pt]{4.818pt}{0.400pt}}
\put(666.0,274.0){\rule[-0.200pt]{4.818pt}{0.400pt}}
\sbox{\plotpoint}{\rule[-0.500pt]{1.000pt}{1.000pt}}%
\put(220,540){\usebox{\plotpoint}}
\put(220.00,540.00){\usebox{\plotpoint}}
\multiput(226,540)(20.756,0.000){0}{\usebox{\plotpoint}}
\multiput(232,540)(20.756,0.000){0}{\usebox{\plotpoint}}
\put(240.76,540.00){\usebox{\plotpoint}}
\multiput(245,540)(20.756,0.000){0}{\usebox{\plotpoint}}
\multiput(251,540)(20.756,0.000){0}{\usebox{\plotpoint}}
\put(261.51,540.00){\usebox{\plotpoint}}
\multiput(264,540)(20.756,0.000){0}{\usebox{\plotpoint}}
\multiput(270,540)(20.756,0.000){0}{\usebox{\plotpoint}}
\multiput(276,540)(20.756,0.000){0}{\usebox{\plotpoint}}
\put(282.27,540.00){\usebox{\plotpoint}}
\multiput(288,540)(20.756,0.000){0}{\usebox{\plotpoint}}
\multiput(295,540)(20.756,0.000){0}{\usebox{\plotpoint}}
\put(303.02,540.00){\usebox{\plotpoint}}
\multiput(307,540)(20.756,0.000){0}{\usebox{\plotpoint}}
\multiput(313,540)(20.756,0.000){0}{\usebox{\plotpoint}}
\put(323.78,540.00){\usebox{\plotpoint}}
\multiput(326,540)(20.756,0.000){0}{\usebox{\plotpoint}}
\multiput(332,540)(20.756,0.000){0}{\usebox{\plotpoint}}
\multiput(338,540)(20.756,0.000){0}{\usebox{\plotpoint}}
\put(344.53,540.00){\usebox{\plotpoint}}
\multiput(351,540)(20.756,0.000){0}{\usebox{\plotpoint}}
\multiput(357,540)(20.756,0.000){0}{\usebox{\plotpoint}}
\put(365.29,540.00){\usebox{\plotpoint}}
\multiput(369,540)(20.756,0.000){0}{\usebox{\plotpoint}}
\multiput(376,540)(20.756,0.000){0}{\usebox{\plotpoint}}
\put(386.04,540.00){\usebox{\plotpoint}}
\multiput(388,540)(20.756,0.000){0}{\usebox{\plotpoint}}
\multiput(394,540)(20.756,0.000){0}{\usebox{\plotpoint}}
\put(406.80,540.00){\usebox{\plotpoint}}
\multiput(407,540)(20.473,-3.412){0}{\usebox{\plotpoint}}
\multiput(413,539)(20.756,0.000){0}{\usebox{\plotpoint}}
\multiput(419,539)(20.756,0.000){0}{\usebox{\plotpoint}}
\put(427.47,539.00){\usebox{\plotpoint}}
\multiput(432,539)(20.756,0.000){0}{\usebox{\plotpoint}}
\multiput(438,539)(20.756,0.000){0}{\usebox{\plotpoint}}
\put(448.23,539.00){\usebox{\plotpoint}}
\multiput(450,539)(20.473,-3.412){0}{\usebox{\plotpoint}}
\multiput(456,538)(20.756,0.000){0}{\usebox{\plotpoint}}
\put(468.90,538.00){\usebox{\plotpoint}}
\multiput(469,538)(20.756,0.000){0}{\usebox{\plotpoint}}
\multiput(475,538)(20.756,0.000){0}{\usebox{\plotpoint}}
\multiput(481,538)(20.547,-2.935){0}{\usebox{\plotpoint}}
\put(489.58,537.00){\usebox{\plotpoint}}
\multiput(494,537)(20.756,0.000){0}{\usebox{\plotpoint}}
\multiput(500,537)(20.473,-3.412){0}{\usebox{\plotpoint}}
\put(510.26,536.00){\usebox{\plotpoint}}
\multiput(512,536)(20.756,0.000){0}{\usebox{\plotpoint}}
\multiput(519,536)(20.473,-3.412){0}{\usebox{\plotpoint}}
\put(530.93,535.00){\usebox{\plotpoint}}
\multiput(531,535)(20.473,-3.412){0}{\usebox{\plotpoint}}
\multiput(537,534)(20.756,0.000){0}{\usebox{\plotpoint}}
\multiput(544,534)(20.473,-3.412){0}{\usebox{\plotpoint}}
\put(551.52,533.00){\usebox{\plotpoint}}
\multiput(556,533)(20.473,-3.412){0}{\usebox{\plotpoint}}
\multiput(562,532)(20.756,0.000){0}{\usebox{\plotpoint}}
\put(572.15,531.41){\usebox{\plotpoint}}
\multiput(575,531)(20.473,-3.412){0}{\usebox{\plotpoint}}
\multiput(581,530)(20.756,0.000){0}{\usebox{\plotpoint}}
\put(592.72,529.05){\usebox{\plotpoint}}
\multiput(593,529)(20.547,-2.935){0}{\usebox{\plotpoint}}
\multiput(600,528)(20.473,-3.412){0}{\usebox{\plotpoint}}
\multiput(606,527)(20.473,-3.412){0}{\usebox{\plotpoint}}
\put(613.21,525.80){\usebox{\plotpoint}}
\multiput(618,525)(20.756,0.000){0}{\usebox{\plotpoint}}
\multiput(624,525)(20.547,-2.935){0}{\usebox{\plotpoint}}
\put(633.69,523.10){\usebox{\plotpoint}}
\multiput(637,522)(20.473,-3.412){0}{\usebox{\plotpoint}}
\multiput(643,521)(20.473,-3.412){0}{\usebox{\plotpoint}}
\put(654.05,519.28){\usebox{\plotpoint}}
\multiput(656,519)(20.473,-3.412){0}{\usebox{\plotpoint}}
\multiput(662,518)(20.473,-3.412){0}{\usebox{\plotpoint}}
\multiput(668,517)(19.690,-6.563){0}{\usebox{\plotpoint}}
\put(674.29,514.95){\usebox{\plotpoint}}
\multiput(680,514)(19.957,-5.702){0}{\usebox{\plotpoint}}
\multiput(687,512)(20.473,-3.412){0}{\usebox{\plotpoint}}
\put(694.52,510.49){\usebox{\plotpoint}}
\multiput(699,509)(20.473,-3.412){0}{\usebox{\plotpoint}}
\multiput(705,508)(19.957,-5.702){0}{\usebox{\plotpoint}}
\put(714.53,505.16){\usebox{\plotpoint}}
\multiput(718,504)(19.690,-6.563){0}{\usebox{\plotpoint}}
\multiput(724,502)(19.690,-6.563){0}{\usebox{\plotpoint}}
\put(734.22,498.59){\usebox{\plotpoint}}
\multiput(736,498)(19.957,-5.702){0}{\usebox{\plotpoint}}
\multiput(743,496)(19.690,-6.563){0}{\usebox{\plotpoint}}
\put(754.01,492.33){\usebox{\plotpoint}}
\multiput(755,492)(19.690,-6.563){0}{\usebox{\plotpoint}}
\multiput(761,490)(19.077,-8.176){0}{\usebox{\plotpoint}}
\put(773.47,485.18){\usebox{\plotpoint}}
\multiput(774,485)(18.564,-9.282){0}{\usebox{\plotpoint}}
\multiput(780,482)(19.690,-6.563){0}{\usebox{\plotpoint}}
\multiput(786,480)(18.564,-9.282){0}{\usebox{\plotpoint}}
\put(792.42,476.82){\usebox{\plotpoint}}
\multiput(799,474)(18.564,-9.282){0}{\usebox{\plotpoint}}
\multiput(805,471)(18.564,-9.282){0}{\usebox{\plotpoint}}
\put(811.16,467.92){\usebox{\plotpoint}}
\multiput(817,465)(19.077,-8.176){0}{\usebox{\plotpoint}}
\put(829.92,459.04){\usebox{\plotpoint}}
\multiput(830,459)(17.270,-11.513){0}{\usebox{\plotpoint}}
\put(836,455){\usebox{\plotpoint}}
\put(220,574){\usebox{\plotpoint}}
\put(220.00,574.00){\usebox{\plotpoint}}
\multiput(226,574)(20.756,0.000){0}{\usebox{\plotpoint}}
\multiput(232,574)(20.756,0.000){0}{\usebox{\plotpoint}}
\put(240.76,574.00){\usebox{\plotpoint}}
\multiput(245,574)(20.756,0.000){0}{\usebox{\plotpoint}}
\multiput(251,574)(20.756,0.000){0}{\usebox{\plotpoint}}
\put(261.51,574.00){\usebox{\plotpoint}}
\multiput(264,574)(20.756,0.000){0}{\usebox{\plotpoint}}
\multiput(270,574)(20.756,0.000){0}{\usebox{\plotpoint}}
\multiput(276,574)(20.756,0.000){0}{\usebox{\plotpoint}}
\put(282.27,574.00){\usebox{\plotpoint}}
\multiput(288,574)(20.756,0.000){0}{\usebox{\plotpoint}}
\multiput(295,574)(20.756,0.000){0}{\usebox{\plotpoint}}
\put(303.02,574.00){\usebox{\plotpoint}}
\multiput(307,574)(20.756,0.000){0}{\usebox{\plotpoint}}
\multiput(313,574)(20.756,0.000){0}{\usebox{\plotpoint}}
\put(323.78,574.00){\usebox{\plotpoint}}
\multiput(326,574)(20.756,0.000){0}{\usebox{\plotpoint}}
\multiput(332,574)(20.473,3.412){0}{\usebox{\plotpoint}}
\multiput(338,575)(20.756,0.000){0}{\usebox{\plotpoint}}
\put(344.45,575.00){\usebox{\plotpoint}}
\multiput(351,575)(20.756,0.000){0}{\usebox{\plotpoint}}
\multiput(357,575)(20.756,0.000){0}{\usebox{\plotpoint}}
\put(365.21,575.00){\usebox{\plotpoint}}
\multiput(369,575)(20.547,2.935){0}{\usebox{\plotpoint}}
\multiput(376,576)(20.756,0.000){0}{\usebox{\plotpoint}}
\put(385.89,576.00){\usebox{\plotpoint}}
\multiput(388,576)(20.473,3.412){0}{\usebox{\plotpoint}}
\multiput(394,577)(20.756,0.000){0}{\usebox{\plotpoint}}
\put(406.56,577.00){\usebox{\plotpoint}}
\multiput(407,577)(20.473,3.412){0}{\usebox{\plotpoint}}
\multiput(413,578)(20.756,0.000){0}{\usebox{\plotpoint}}
\multiput(419,578)(20.473,3.412){0}{\usebox{\plotpoint}}
\put(427.15,579.00){\usebox{\plotpoint}}
\multiput(432,579)(20.473,3.412){0}{\usebox{\plotpoint}}
\multiput(438,580)(20.473,3.412){0}{\usebox{\plotpoint}}
\put(447.69,581.62){\usebox{\plotpoint}}
\multiput(450,582)(20.473,3.412){0}{\usebox{\plotpoint}}
\multiput(456,583)(20.547,2.935){0}{\usebox{\plotpoint}}
\put(468.19,584.87){\usebox{\plotpoint}}
\multiput(469,585)(20.473,3.412){0}{\usebox{\plotpoint}}
\multiput(475,586)(20.473,3.412){0}{\usebox{\plotpoint}}
\multiput(481,587)(20.547,2.935){0}{\usebox{\plotpoint}}
\put(488.69,588.11){\usebox{\plotpoint}}
\multiput(494,589)(19.690,6.563){0}{\usebox{\plotpoint}}
\multiput(500,591)(20.473,3.412){0}{\usebox{\plotpoint}}
\put(508.81,592.94){\usebox{\plotpoint}}
\multiput(512,594)(19.957,5.702){0}{\usebox{\plotpoint}}
\multiput(519,596)(19.690,6.563){0}{\usebox{\plotpoint}}
\put(528.60,599.20){\usebox{\plotpoint}}
\multiput(531,600)(19.690,6.563){0}{\usebox{\plotpoint}}
\multiput(537,602)(19.957,5.702){0}{\usebox{\plotpoint}}
\put(548.38,605.46){\usebox{\plotpoint}}
\multiput(550,606)(18.564,9.282){0}{\usebox{\plotpoint}}
\multiput(556,609)(18.564,9.282){0}{\usebox{\plotpoint}}
\put(567.04,614.52){\usebox{\plotpoint}}
\multiput(568,615)(19.957,5.702){0}{\usebox{\plotpoint}}
\multiput(575,617)(17.270,11.513){0}{\usebox{\plotpoint}}
\put(585.64,623.32){\usebox{\plotpoint}}
\multiput(587,624)(18.564,9.282){0}{\usebox{\plotpoint}}
\multiput(593,627)(18.021,10.298){0}{\usebox{\plotpoint}}
\put(603.71,633.48){\usebox{\plotpoint}}
\multiput(606,635)(17.270,11.513){0}{\usebox{\plotpoint}}
\multiput(612,639)(17.270,11.513){0}{\usebox{\plotpoint}}
\put(620.75,645.30){\usebox{\plotpoint}}
\multiput(624,648)(18.021,10.298){0}{\usebox{\plotpoint}}
\multiput(631,652)(15.945,13.287){0}{\usebox{\plotpoint}}
\put(637.51,657.42){\usebox{\plotpoint}}
\multiput(643,662)(15.945,13.287){0}{\usebox{\plotpoint}}
\put(653.40,670.77){\usebox{\plotpoint}}
\multiput(656,673)(14.676,14.676){0}{\usebox{\plotpoint}}
\multiput(662,679)(14.676,14.676){0}{\usebox{\plotpoint}}
\put(668.25,685.25){\usebox{\plotpoint}}
\multiput(674,691)(13.508,15.759){0}{\usebox{\plotpoint}}
\put(682.59,700.22){\usebox{\plotpoint}}
\multiput(687,704)(13.508,15.759){0}{\usebox{\plotpoint}}
\put(696.44,715.58){\usebox{\plotpoint}}
\multiput(699,719)(12.453,16.604){0}{\usebox{\plotpoint}}
\put(709.27,731.88){\usebox{\plotpoint}}
\multiput(712,735)(12.064,16.889){0}{\usebox{\plotpoint}}
\put(717,742){\usebox{\plotpoint}}
\put(220,392){\usebox{\plotpoint}}
\put(220.00,392.00){\usebox{\plotpoint}}
\multiput(226,392)(20.756,0.000){0}{\usebox{\plotpoint}}
\multiput(232,392)(20.756,0.000){0}{\usebox{\plotpoint}}
\put(240.76,392.00){\usebox{\plotpoint}}
\multiput(245,392)(20.756,0.000){0}{\usebox{\plotpoint}}
\multiput(251,392)(20.756,0.000){0}{\usebox{\plotpoint}}
\put(261.51,392.00){\usebox{\plotpoint}}
\multiput(264,392)(20.473,-3.412){0}{\usebox{\plotpoint}}
\multiput(270,391)(20.756,0.000){0}{\usebox{\plotpoint}}
\multiput(276,391)(20.756,0.000){0}{\usebox{\plotpoint}}
\put(282.18,391.00){\usebox{\plotpoint}}
\multiput(288,391)(20.756,0.000){0}{\usebox{\plotpoint}}
\multiput(295,391)(20.473,-3.412){0}{\usebox{\plotpoint}}
\put(302.86,390.00){\usebox{\plotpoint}}
\multiput(307,390)(20.756,0.000){0}{\usebox{\plotpoint}}
\multiput(313,390)(20.756,0.000){0}{\usebox{\plotpoint}}
\put(323.56,389.41){\usebox{\plotpoint}}
\multiput(326,389)(20.756,0.000){0}{\usebox{\plotpoint}}
\multiput(332,389)(20.756,0.000){0}{\usebox{\plotpoint}}
\multiput(338,389)(20.473,-3.412){0}{\usebox{\plotpoint}}
\put(344.20,388.00){\usebox{\plotpoint}}
\multiput(351,388)(20.473,-3.412){0}{\usebox{\plotpoint}}
\multiput(357,387)(20.756,0.000){0}{\usebox{\plotpoint}}
\put(364.87,387.00){\usebox{\plotpoint}}
\multiput(369,387)(20.547,-2.935){0}{\usebox{\plotpoint}}
\multiput(376,386)(20.756,0.000){0}{\usebox{\plotpoint}}
\put(385.51,385.41){\usebox{\plotpoint}}
\multiput(388,385)(20.756,0.000){0}{\usebox{\plotpoint}}
\multiput(394,385)(20.473,-3.412){0}{\usebox{\plotpoint}}
\put(406.09,383.13){\usebox{\plotpoint}}
\multiput(407,383)(20.756,0.000){0}{\usebox{\plotpoint}}
\multiput(413,383)(20.473,-3.412){0}{\usebox{\plotpoint}}
\multiput(419,382)(20.756,0.000){0}{\usebox{\plotpoint}}
\put(426.73,381.75){\usebox{\plotpoint}}
\multiput(432,381)(20.473,-3.412){0}{\usebox{\plotpoint}}
\multiput(438,380)(20.756,0.000){0}{\usebox{\plotpoint}}
\put(447.31,379.45){\usebox{\plotpoint}}
\multiput(450,379)(20.473,-3.412){0}{\usebox{\plotpoint}}
\multiput(456,378)(20.756,0.000){0}{\usebox{\plotpoint}}
\put(467.88,377.19){\usebox{\plotpoint}}
\multiput(469,377)(20.473,-3.412){0}{\usebox{\plotpoint}}
\multiput(475,376)(20.473,-3.412){0}{\usebox{\plotpoint}}
\multiput(481,375)(20.547,-2.935){0}{\usebox{\plotpoint}}
\put(488.38,374.00){\usebox{\plotpoint}}
\multiput(494,374)(20.473,-3.412){0}{\usebox{\plotpoint}}
\multiput(500,373)(20.473,-3.412){0}{\usebox{\plotpoint}}
\put(508.93,371.51){\usebox{\plotpoint}}
\multiput(512,371)(20.547,-2.935){0}{\usebox{\plotpoint}}
\multiput(519,370)(20.473,-3.412){0}{\usebox{\plotpoint}}
\put(529.43,368.26){\usebox{\plotpoint}}
\multiput(531,368)(20.473,-3.412){0}{\usebox{\plotpoint}}
\multiput(537,367)(20.547,-2.935){0}{\usebox{\plotpoint}}
\put(549.92,365.01){\usebox{\plotpoint}}
\multiput(550,365)(20.473,-3.412){0}{\usebox{\plotpoint}}
\multiput(556,364)(20.473,-3.412){0}{\usebox{\plotpoint}}
\multiput(562,363)(20.473,-3.412){0}{\usebox{\plotpoint}}
\put(570.41,361.66){\usebox{\plotpoint}}
\multiput(575,361)(20.473,-3.412){0}{\usebox{\plotpoint}}
\multiput(581,360)(20.473,-3.412){0}{\usebox{\plotpoint}}
\put(590.90,358.35){\usebox{\plotpoint}}
\multiput(593,358)(20.547,-2.935){0}{\usebox{\plotpoint}}
\multiput(600,357)(19.690,-6.563){0}{\usebox{\plotpoint}}
\put(611.16,354.14){\usebox{\plotpoint}}
\multiput(612,354)(20.473,-3.412){0}{\usebox{\plotpoint}}
\multiput(618,353)(20.473,-3.412){0}{\usebox{\plotpoint}}
\multiput(624,352)(20.547,-2.935){0}{\usebox{\plotpoint}}
\put(631.63,350.79){\usebox{\plotpoint}}
\multiput(637,349)(20.473,-3.412){0}{\usebox{\plotpoint}}
\multiput(643,348)(20.473,-3.412){0}{\usebox{\plotpoint}}
\put(651.82,346.20){\usebox{\plotpoint}}
\multiput(656,345)(20.473,-3.412){0}{\usebox{\plotpoint}}
\multiput(662,344)(20.473,-3.412){0}{\usebox{\plotpoint}}
\put(672.02,341.66){\usebox{\plotpoint}}
\multiput(674,341)(20.473,-3.412){0}{\usebox{\plotpoint}}
\multiput(680,340)(20.547,-2.935){0}{\usebox{\plotpoint}}
\put(692.23,337.26){\usebox{\plotpoint}}
\multiput(693,337)(20.473,-3.412){0}{\usebox{\plotpoint}}
\multiput(699,336)(19.690,-6.563){0}{\usebox{\plotpoint}}
\multiput(705,334)(20.547,-2.935){0}{\usebox{\plotpoint}}
\put(712.44,332.85){\usebox{\plotpoint}}
\multiput(718,331)(20.473,-3.412){0}{\usebox{\plotpoint}}
\multiput(724,330)(19.690,-6.563){0}{\usebox{\plotpoint}}
\put(732.46,327.59){\usebox{\plotpoint}}
\multiput(736,327)(19.957,-5.702){0}{\usebox{\plotpoint}}
\multiput(743,325)(19.690,-6.563){0}{\usebox{\plotpoint}}
\put(752.51,322.41){\usebox{\plotpoint}}
\multiput(755,322)(19.690,-6.563){0}{\usebox{\plotpoint}}
\multiput(761,320)(19.957,-5.702){0}{\usebox{\plotpoint}}
\put(772.56,317.24){\usebox{\plotpoint}}
\multiput(774,317)(19.690,-6.563){0}{\usebox{\plotpoint}}
\multiput(780,315)(19.690,-6.563){0}{\usebox{\plotpoint}}
\multiput(786,313)(20.473,-3.412){0}{\usebox{\plotpoint}}
\put(792.55,311.84){\usebox{\plotpoint}}
\multiput(799,310)(19.690,-6.563){0}{\usebox{\plotpoint}}
\multiput(805,308)(19.690,-6.563){0}{\usebox{\plotpoint}}
\put(812.32,305.56){\usebox{\plotpoint}}
\multiput(817,304)(20.547,-2.935){0}{\usebox{\plotpoint}}
\multiput(824,303)(19.690,-6.563){0}{\usebox{\plotpoint}}
\put(832.31,300.23){\usebox{\plotpoint}}
\put(836,299){\usebox{\plotpoint}}
\put(220,296){\usebox{\plotpoint}}
\put(220.00,296.00){\usebox{\plotpoint}}
\multiput(226,296)(20.756,0.000){0}{\usebox{\plotpoint}}
\multiput(232,296)(20.756,0.000){0}{\usebox{\plotpoint}}
\put(240.76,296.00){\usebox{\plotpoint}}
\multiput(245,296)(20.756,0.000){0}{\usebox{\plotpoint}}
\multiput(251,296)(20.756,0.000){0}{\usebox{\plotpoint}}
\put(261.51,296.00){\usebox{\plotpoint}}
\multiput(264,296)(20.756,0.000){0}{\usebox{\plotpoint}}
\multiput(270,296)(20.756,0.000){0}{\usebox{\plotpoint}}
\multiput(276,296)(20.756,0.000){0}{\usebox{\plotpoint}}
\put(282.27,296.00){\usebox{\plotpoint}}
\multiput(288,296)(20.547,-2.935){0}{\usebox{\plotpoint}}
\multiput(295,295)(20.756,0.000){0}{\usebox{\plotpoint}}
\put(302.95,295.00){\usebox{\plotpoint}}
\multiput(307,295)(20.756,0.000){0}{\usebox{\plotpoint}}
\multiput(313,295)(20.756,0.000){0}{\usebox{\plotpoint}}
\put(323.71,295.00){\usebox{\plotpoint}}
\multiput(326,295)(20.473,-3.412){0}{\usebox{\plotpoint}}
\multiput(332,294)(20.756,0.000){0}{\usebox{\plotpoint}}
\multiput(338,294)(20.756,0.000){0}{\usebox{\plotpoint}}
\put(344.38,294.00){\usebox{\plotpoint}}
\multiput(351,294)(20.473,-3.412){0}{\usebox{\plotpoint}}
\multiput(357,293)(20.756,0.000){0}{\usebox{\plotpoint}}
\put(365.05,293.00){\usebox{\plotpoint}}
\multiput(369,293)(20.547,-2.935){0}{\usebox{\plotpoint}}
\multiput(376,292)(20.756,0.000){0}{\usebox{\plotpoint}}
\put(385.74,292.00){\usebox{\plotpoint}}
\multiput(388,292)(20.473,-3.412){0}{\usebox{\plotpoint}}
\multiput(394,291)(20.756,0.000){0}{\usebox{\plotpoint}}
\put(406.41,291.00){\usebox{\plotpoint}}
\multiput(407,291)(20.473,-3.412){0}{\usebox{\plotpoint}}
\multiput(413,290)(20.756,0.000){0}{\usebox{\plotpoint}}
\multiput(419,290)(20.756,0.000){0}{\usebox{\plotpoint}}
\put(427.06,289.71){\usebox{\plotpoint}}
\multiput(432,289)(20.756,0.000){0}{\usebox{\plotpoint}}
\multiput(438,289)(20.473,-3.412){0}{\usebox{\plotpoint}}
\put(447.68,288.00){\usebox{\plotpoint}}
\multiput(450,288)(20.473,-3.412){0}{\usebox{\plotpoint}}
\multiput(456,287)(20.756,0.000){0}{\usebox{\plotpoint}}
\put(468.28,286.12){\usebox{\plotpoint}}
\multiput(469,286)(20.756,0.000){0}{\usebox{\plotpoint}}
\multiput(475,286)(20.473,-3.412){0}{\usebox{\plotpoint}}
\multiput(481,285)(20.756,0.000){0}{\usebox{\plotpoint}}
\put(488.93,284.84){\usebox{\plotpoint}}
\multiput(494,284)(20.756,0.000){0}{\usebox{\plotpoint}}
\multiput(500,284)(20.473,-3.412){0}{\usebox{\plotpoint}}
\put(509.54,283.00){\usebox{\plotpoint}}
\multiput(512,283)(20.547,-2.935){0}{\usebox{\plotpoint}}
\multiput(519,282)(20.473,-3.412){0}{\usebox{\plotpoint}}
\put(530.14,281.00){\usebox{\plotpoint}}
\multiput(531,281)(20.473,-3.412){0}{\usebox{\plotpoint}}
\multiput(537,280)(20.756,0.000){0}{\usebox{\plotpoint}}
\multiput(544,280)(20.473,-3.412){0}{\usebox{\plotpoint}}
\put(550.72,278.88){\usebox{\plotpoint}}
\multiput(556,278)(20.756,0.000){0}{\usebox{\plotpoint}}
\multiput(562,278)(20.473,-3.412){0}{\usebox{\plotpoint}}
\put(571.28,276.53){\usebox{\plotpoint}}
\multiput(575,276)(20.473,-3.412){0}{\usebox{\plotpoint}}
\multiput(581,275)(20.756,0.000){0}{\usebox{\plotpoint}}
\put(591.85,274.19){\usebox{\plotpoint}}
\multiput(593,274)(20.547,-2.935){0}{\usebox{\plotpoint}}
\multiput(600,273)(20.473,-3.412){0}{\usebox{\plotpoint}}
\multiput(606,272)(20.756,0.000){0}{\usebox{\plotpoint}}
\put(612.43,271.93){\usebox{\plotpoint}}
\multiput(618,271)(20.473,-3.412){0}{\usebox{\plotpoint}}
\multiput(624,270)(20.547,-2.935){0}{\usebox{\plotpoint}}
\put(632.93,268.68){\usebox{\plotpoint}}
\multiput(637,268)(20.756,0.000){0}{\usebox{\plotpoint}}
\multiput(643,268)(20.473,-3.412){0}{\usebox{\plotpoint}}
\put(653.50,266.36){\usebox{\plotpoint}}
\multiput(656,266)(20.473,-3.412){0}{\usebox{\plotpoint}}
\multiput(662,265)(20.473,-3.412){0}{\usebox{\plotpoint}}
\put(673.98,263.00){\usebox{\plotpoint}}
\multiput(674,263)(20.473,-3.412){0}{\usebox{\plotpoint}}
\multiput(680,262)(20.547,-2.935){0}{\usebox{\plotpoint}}
\multiput(687,261)(20.473,-3.412){0}{\usebox{\plotpoint}}
\put(694.50,260.00){\usebox{\plotpoint}}
\multiput(699,260)(20.473,-3.412){0}{\usebox{\plotpoint}}
\multiput(705,259)(20.547,-2.935){0}{\usebox{\plotpoint}}
\put(715.06,257.49){\usebox{\plotpoint}}
\multiput(718,257)(20.473,-3.412){0}{\usebox{\plotpoint}}
\multiput(724,256)(20.473,-3.412){0}{\usebox{\plotpoint}}
\put(735.53,254.08){\usebox{\plotpoint}}
\multiput(736,254)(20.547,-2.935){0}{\usebox{\plotpoint}}
\multiput(743,253)(19.690,-6.563){0}{\usebox{\plotpoint}}
\multiput(749,251)(20.473,-3.412){0}{\usebox{\plotpoint}}
\put(755.79,249.87){\usebox{\plotpoint}}
\multiput(761,249)(20.547,-2.935){0}{\usebox{\plotpoint}}
\multiput(768,248)(20.473,-3.412){0}{\usebox{\plotpoint}}
\put(776.29,246.62){\usebox{\plotpoint}}
\multiput(780,246)(20.473,-3.412){0}{\usebox{\plotpoint}}
\multiput(786,245)(20.473,-3.412){0}{\usebox{\plotpoint}}
\put(796.78,243.32){\usebox{\plotpoint}}
\multiput(799,243)(19.690,-6.563){0}{\usebox{\plotpoint}}
\multiput(805,241)(20.473,-3.412){0}{\usebox{\plotpoint}}
\multiput(811,240)(20.473,-3.412){0}{\usebox{\plotpoint}}
\put(817.03,239.00){\usebox{\plotpoint}}
\multiput(824,238)(20.473,-3.412){0}{\usebox{\plotpoint}}
\multiput(830,237)(19.690,-6.563){0}{\usebox{\plotpoint}}
\put(836,235){\usebox{\plotpoint}}
\end{picture}

\end{center}
\vspace{-5ex}
\caption{ Comparison of speed of light squared \eqn{csq} for D234c and SW
actions. Pseudoscalar (P) and Vector (V) mesons at $P/V\approx 0.7$ are shown.
}
\label{fig:csq}
\vspace{-3ex}
\end{figure}

\section{Lorentz Violations}
From Figure~\ref{fig:phi70} the SW action appears to work well even
at~0.4\,fm, however it seriously violates Lorentz invariance. To see
this, consider the speed of light
\beq\label{csq}
c^2(\pv) = \frac{E^2(\pv)-E^2(0)}{\pv^2}
\eeq
which should equal~1, for all~$\pv$. 
This quantity is particularly sensistive to the $C_3$~term
in the D234c action since this term 
cancels the leading Lorentz-violating error.
Our results for $c^2$, for both pseudoscalar and vector
mesons, are shown in Figure~\ref{fig:csq}. At 0.4~\fm, D234c is
dramatically superior: it deviates from $c^2= 1$ by only 3--5\% at
zero momentum, and by less than 10\% even at momenta of order $1.5/a$,
while SW gives results that deviate by 40--60\% or more for all
momenta, including zero. 
The reason for SW's poor performance is
that the strange quark is relatively
massive at our lattice spacings: the $\phi$ mass
is $2.1/a$  at $a=0.4~\fm$.  These results illustrate that D234c
is far more accurate for hadrons with large masses.

The $C_3$ term is the only $a^2$ correction that breaks Lorentz
invariance, so we can use $c^2$ to tune $C_3$~nonperturbatively. At
0.4\,fm\ we tuned $C_3$ to make the dispersion relation for
the lightest meson, the pseudoscalar, perfect at low momentum: when
\beq
C_3 = 1.2 \approx 1 + \alpha_s/2
\eeq
we obtain the $c^2(p\=0)$ results shown in table~\ref{tab:csq}. The
pseudoscalar (P) moves to within $\pm2\%$ of $c^2=1$.  (The vector (V)
shows 20\% deviations, because it is 25\% heavier than the P at this
quark mass; lowering the quark mass to $P/V=0.6$ we see that,
within errors, both P
and V show no Lorentz violation at low momentum.)

As with $C_F$, the non-perturbative estimate of the quantum corrections
to the mean-link TI tree-level value of
$C_3$ indicates that they are perturbative in size, and
a perturbative calculation of $C_3$ is currently in progress.

\def\st{\rule[-1.5ex]{0em}{4ex}} % Strut to give correct row spacing in tables
\begin{table}[t]
% space before first and after last column: 1.5pc
% space between columns: 3.0pc (twice the above)
%\setlength{\tabcolsep}{1.5pc}
% -----------------------------------------------------
% adapted from TeX book, p. 241
%\newlength{\digitwidth}
\settowidth{\digitwidth}{\rm 0}
\catcode`?=\active \def?{\kern\digitwidth}
% -----------------------------------------------------
%\caption{Biologically treated effluents (mg/l)}
%\begin{tabular*}{\textwidth}{llllll}
\begin{tabular}{llll}
\hline
\st  & \multicolumn{2}{c}{$P/V\approx 0.76$} & $P/V\approx 0.6$ \\
\st & $C_3=1.0$ & $C_3=1.2$ & $C_3=1.2$ \\
\hline
\st P & 0.95(2) & 1.00(1) & 0.99(2) \\
\st V & 1.02(4) & 1.21(5) & 1.04(3) \\
\hline
\vspace{-0.5ex}
\end{tabular}
\caption{
$c^2({\bf p})$ extrapolated to $p=0$.
Tuning P to $c^2=1$ gives $C_3=1.2$.
Only at lower quark mass ($P/V=0.6$) can both P and V
be tuned simultaneously.
}
\label{tab:csq}
\vspace{-5ex}
\end{table}

\def\order{{\cal O}}
\section{Conclusions}

The D234c action is 
a very useful tool for lattice
calculations involving large quark masses or momenta,
making it particularly suitable for coarse lattices,
high-momentum form factors, heavy quarks, and finite temperature
studies. With more precise (perturbative or non-perturbative) tuning
of the leading coefficients it should be accurate to within a few
percent at lattice spacings as large as~0.4~\fm\ and meson masses as
large as $2/a\!\approx\!1~\GeV$.  We are also continuing our previous
work with D234 on anisotropic lattices, which will push the range of
accessible energies even higher. 

\vspace{1ex}\noindent
We acknowledge the Cornell Theory Center for the use of its SP-2
computer in this research. MGA supported by DOE DE-FG02-90ER40542.

\end{document}